\pdfoutput=1

\documentclass[pra,twocolumn,10pt,superscriptaddress, footnoteinbib]{revtex4}
\usepackage{amsmath}
\usepackage{latexsym}
\usepackage{amssymb}
\usepackage{graphics,epstopdf}
\usepackage[colorlinks=true, citecolor=blue, urlcolor=blue ]{hyperref}
\usepackage{epsf,graphics,graphicx}

\newcommand{\n}{\noindent}

\newcommand{\ed}{\end{document}}
\newcommand{\beq}{\begin{equation}}
\newcommand{\eeq}{\end{equation}}

\begin{document}
\title{Relativistic electron vortex beams in a laser field}
\author{ Pratul Bandyopadhyay \footnote{Electronic address:{b$_{-}$pratul@yahoo.co.in}}, Banasri Basu\footnote{Electronic address:{sribbasu@gmail.com}}${}^{}$, Debashree Chowdhury\footnote{Electronic
address:{debashreephys@gmail.com}}${}^{}$  }
\affiliation{Physics and
Applied Mathematics Unit, Indian Statistical Institute,\\
 203
Barrackpore Trunk Road, Kolkata 700 108, India}


\begin{abstract}
\n
The orbital angular momentum Hall effect and spin Hall effect of electron vortex beams (EVB) have been studied for the EVBs interacting with laser field. In the scenario of paraxial beam, the cumulative effect of the orbit-orbit interaction of EVBs and laser fields  drives the orbital Hall effect,  which in turn produces  a shift of the center of the beam from that of the field-free case towards the polarization axis of photons. Besides, for non-paraxial beams one can also perceive a similar shift of the center of the beam  owing to spin Hall effect involving spin-orbit interaction. Our analysis suggests that the shift in the paraxial beams will always be larger than that in non-paraxial beams.

\end{abstract}

\maketitle

The experimental demonstration \cite{1} for the production of electron vortex beam (EVB) 
with an orbital angular momentum (OAM) projection of upto 200$\hbar$ and energy $ \sim$ 200-300 KeV have upsurged the interest of theoretical physicists to work in this area  which was started from the theoretical prediction of EVB \cite{bliokh07}.
In free EVBs both the spin angular momentum (SAM) and the OAM of an electron give rise to an intrinsic spin-orbit interaction (SOI) \cite{bliokh 2011} which is also a topic of recent attraction. The recent theoretical investigation of the interaction of relativistic electron vortex beams with laser light \cite{laser} is also of much importance from the aspect of light matter interaction. The exact analytical solutions, obtained when the Dirac-Volkov wave function is used to describe the monoenergetic distribution of electrons with well defined OAM in vortex beams,  explicitly shows that the OAM components of the laser field couples to the total angular momentum (TAM) of the electron and the center of the beam is shifted along the polarization direction of the laser field with respect to the center of the field-free EVB \cite{laser}. In their analysis they have used the calculation of the probability density of finding an electron in the beam profile. 
This enthralling paper \cite{laser} motivated us to probe the physical mechanism responsible for the above mentioned shift of the center of the beam. 
 Very recently, we have studied  \cite{our,evb} the dynamics of field free EVBs and EVBs in a time dependent magnetic field that revealed  how the spin Hall effect and the spin filter configuration of EVBs arise. In this formulation, we have utilized the role of the geometric phase acquired by the scalar electron orbiting around the vortex line in the geometrodynamics of EVBS. It is worth studying
 the situation when EVBs interact with a laser field in this framework. 


In this letter, we explore the dynamics of relativistic electron vortex beams in a laser field from the perspective of geometric phase. To this end, we consider the skyrmionic representation of a fermion, where an electron is depicted as a scalar particle moving around the vortex line, which is topologically equivalent to a magnetic flux line giving rise to the spin degrees of freedom. The geometrodynamics \cite{our,evb} of electron vortex beams is governed by the Berry phase \cite{berry} acquired by the scalar electron moving around the vortex line. This phase term vanishes when the polar angle $\theta$ between the vortex line and the wavefront propagation direction (z axis) is zero. Bessel beam in this case implies that the plane wave wave vector makes an angle $\theta_{0}$ with the $z$ axis such that it depicts the situation in the limiting case $\theta_{0} \longrightarrow 0.$ 
This corresponds to the paraxial regime of the EVB and in this case there is no spin-orbit interaction(SOI). When
such EVBs interact with a laser field, OAM Hall effect will arise inducing a shift of the center of the beam from the field-free case towards the polarization axis of photons. 
On the otherhand, for non-paraxial beams where the polar angle $\theta$ between the vortex line  and the wavefront propagation direction ($z-$ axis) is non-zero, SOI is  switched on and the corresponding Berry phase has a non-zero value.  Bessel beams in this case correspond to the situation when the polar angle $\theta_0$  of the plane wave wave vector with the $z-$ axis takes distinct non-zero value. 
When the polar angle $\theta$ is $\frac{\pi}{2},$ the Berry phase acquired by the scalar electron around the vortex line involves quantized monopole charge $\mu = \frac{1}{2}$. However, for any arbitrary angle $(\theta \neq 0,\frac{\pi}{2}),$ the corresponding Berry phase involves non-quantized monopole charge. In this scenario, Bessel beams involve tilted vortices having an arbitrary non-zero angle of the plane wave wave vector with respect to the wave front propagation direction. It is interesting to note that the propagation of electron vortex beams in free space with tilted vortices give rise to spin Hall effect (SHE) \cite{our}. When such EVBs interact with a laser field, it is found that there will be a shift of the center of the beam towards the polarization axis of photons due to spin Hall effect.

The Dirac equation of an electron coupled to an external electromagnetic field is given by \cite{book}
\beq\left[(\vec{p}-e\vec{A})^{2} - m^{2} - ieF_{\mu\nu}\sigma^{\mu\nu}/2\right]\psi = 0  \label{l1},\eeq
where $\hat{p}_{\mu} = (i\partial_{t}, - i\nabla)$ is the electron four momentum operator and $e$ is the electron charge, $F_{\mu\nu}$ is the electromagnetic field tensor, $2\sigma^{\mu\nu} = \gamma^{\mu}\gamma^{\nu} - \gamma^{\nu}\gamma^{\mu},$ where $\gamma^{\mu}$ are the $4\times 4$ Dirac matrices. For a plane wave field $A_{\mu}(\zeta),$ with $\zeta = kx$ being the laser phase, the exact solution of eqn (\ref{l1}) is of the Dirac-Volkov form \cite{book}
\beq \psi_{p}(x) = \left[1 + \frac{e(\gamma k)(\gamma A)}{2(kp)}\right]\frac{u_{p}}{\sqrt{2E}}e^{iS}\label{l2}\eeq
with $E^{2} = p^{2} + m^{2},$ where $E(p)$ is the energy (momentum) of the electron. Here $S = -(px) - {\cal F} +{\cal G}$  with

\beq {\cal F} = \int_{0}^{\zeta} d\zeta^{~'} \frac{e(pA(\zeta^{~'}))}{(kp)}, ~~ {\cal G} =\int_{0}^{\zeta} d\zeta^{~'} \frac{e^{2}(A^{2}(\zeta^{~'}))}{2(kp)} \label{l3}.\eeq
The spinorial variable $u_{p}$ denotes the positive energy-momentum eigenstate of the Dirac equation in the free space. The spin states of an electron are chosen to be the eigenstates $w^{s} = (\alpha, \beta)^{T}$ of the $\sigma_{z}$ operator with eigenvalues $s_{z} \pm \frac{1}{2}.$ We consider that the electrons and the linearly polarized photons of the external field propagate anti-parallel to each other. The propagation of the electron is chosen to be directed along the $z$ axis so that the laser propagates backward along $z$ which implies $\zeta = \omega t + kz.$ The polarization axis of photons is taken to be in the $y$ direction. For a monoenergetic distribution of the momentum over some cone with $p_{0} = const$ and fixed polar angle $\theta_{0}$ with regard to the propagation axis of the beam we have  
$p_{\parallel 0} = p_{0}cos\theta_{0}$ and $p_{\perp 0} = p_{0}sin\theta_{0}.$ Using cylindrical coordinates in momentum space $\vec{p} = (p_{\perp}, \phi, p_{\parallel}) = (psin\theta,\phi, pcos\theta)$ the Volkov-Bessel solutions can be constructed from the Dirac Volkov solution given by eqn. (2) as
\beq \psi_{l}(x) = \int \tilde{\psi}_{l}(\vec{p})\psi_{p}(x)p_{\perp} dp_{\perp}d\phi,\label{l4}\eeq  
with \beq  \tilde{\psi}_{l}(\vec{p}) = \delta(p_{\perp} - p_{\perp 0})\frac{e^{il\phi}}{2\pi i^{l}p_{\perp 0}}\label{l5}.\eeq
Integrating eqn (\ref{l4}) leads to the Volkov -Bessel state
\beq \psi_{l} (\vec{r},t) = \left[1 + \frac{e}{2(kp_{0})}(\gamma k)(\gamma A)\right]\sum^{+\infty}_{n=-\infty} i^{n} J_{n}(f_{0})\psi_{l+n}(\vec{r},t)\label{l6}\eeq 
with $f_{0} = f(p_{0}),$ where $f = \int_{0}^{\zeta}d\zeta^{~'} ep_{\perp}A(\zeta^{~'})/(kp).$ Taking into account $\xi = p_{\perp 0}r$ the states \cite{laser}
\begin{widetext}
\beq \psi_{l+n}(\vec{r},t) = \frac{e^{i\phi}}{\sqrt{2}}\left[\left(\begin{array}{cr}
   \sqrt{1 + \frac{m}{E_{0}}}w^{s} \\
    \sqrt{1 - \frac{m}{E_{0}}}\sigma_{z}w^{s}cos\theta_{0}

   \end{array}\right)e^{i(l+n)\varphi}J_{l+n}(\xi) +  \left(\begin{array}{ccr}
     0  \\
    0 \\
 {\cal A}

\end{array}\right)e^{i(l+n+1)\varphi}J_{l+n+1}(\xi) -  \left(\begin{array}{ccr}
           0  \\
             0 \\
            {\cal B}
           \end{array}\right)e^{i(l+n-1)\varphi}J_{l+n-1}(\xi)\right]\label{l7}\eeq
           \end{widetext}
with a phase $\phi (z,t) = p_{\parallel 0}z -E_{0}t + {\cal G}(p_{0}),$ spinors ${\cal A} = (0, i\sqrt{\Delta}\alpha)^{T}$ and ${\cal B} = (i\sqrt{\Delta}\beta, 0)^{T}$  and $\Delta = (1 - \frac{m}{E_{0}})sin^{2}\theta_{0}$ appear as Bessel type solution of the Dirac equation with the OAM of an electron in free space modified as $l+n,$ where $n$ is an additional OAM due to laser. 

In the framework of the skyrmionic model of an electron, an internal variable is introduced to represent the direction vector essentially representing a vortex line, giving rise to the spin degrees of freedom \cite{7,8} where the spin appears as an SU(2) gauge bundle. This represents a gauge theoretical extension of the space-time coordinate which can be written as gauge covariant operator acting as functions in phase space \begin{eqnarray}\label{l8}
Q_\mu =  -i\left( \frac{\partial}{\partial p_\mu}+{\cal{A}}_\mu (p)\right) 
\end{eqnarray}   
where ${\cal{A}}_{\mu}(p)$ is the momentum dependent SU(2) gauge field.  Here $p_{\mu}$ denotes the mean momentum of the external observable space. In this formalism, a massive fermion appears as a skyrmion \cite{11,12}. 
The Berry phase acquired by the scalar particle after encircling the closed path around the vortex line, which is equivalent to the magnetic flux line is $2\pi\mu,$ where $\mu$ is the monopole charge associated with the magnetic flux line \cite{19}. When the monopole is located at the origin of a unit sphere, the Berry phase is given by $\phi_{B} = \mu\Omega(C),$ where $\Omega(C)$ is the solid angle subtended by the closed contour at the origin which is given by \beq \Omega(C) = \int_{C}(1 - cos \theta)d\phi = 2\pi(1- cos\theta).\label{22}\eeq
Here $\theta$ is the polar angle of the vortex line with the quantization axis(z axis). So for $\mu = \frac{1}{2},$ we have the phase
\beq \phi_{B} = \pi(1 - cos\theta)\label{23}.\eeq
This corresponds to the flux associated with the monopole passing through the surface spanning the closed contour. Transforming to a reference frame where the scalar electron is considered to be fixed and the vortex state (spin state) moves in the field of the magnetic monopole around a closed path, $\phi_{B}$ in eqn (\ref{23}) corresponds to the geometric phase acquired by the vortex state. The angle $\theta$ represents the deviation of the vortex line from the $z$ axis. Equating this phase $\phi_{B}$ in eqn (\ref{23}) with $2\pi\mu$ which is the geometric phase acquired by the scalar electron moving around the vortex line in a closed path, we find that the effective monopole charge associated with a vortex line having polar angle $\theta$ with the $z$ axis is given by
$\mu = \frac{1}{2}(1 - cos\theta)\label{26}.$ 

This suggests that for $\theta = 0 $ and $\frac{\pi}{2}$, $\mu$ takes quantized values but for $0 < ~\theta<~\frac{\pi}{2}$ it is non-quantized. When the vortex line representing the spin axis is parallel to the wave propagation direction implying $\theta = 0,$ so that the Berry phase vanishes, we have the paraxial vortex beam. 
For
$\theta = \frac{\pi}{2}$ the vortex line is orthogonal to the wavefront propagation direction. For other values of $\theta$, corresponding to non-quantized monopole charge, the vortex line is tilted in an arbitrary direction. This implies the deviation of the spin axis from the $z$ axis and represents the anisotropic feature associated with the system. Bessel beams in this case involve tilted vortices having non-zero arbitrary angle of the plane wave wave vector with respect to the wavefront propagation direction. These three states having $\theta = 0,\frac{\pi}{2}$ and $\neq (0,\frac{\pi}{2})$ correspond to the screw, edge and mixed edge-screw dislocations in optical vortices respectively.

Denoting the spatial coordinate of the electron as $\vec{R},$ we can write from equation (\ref{l8})\cite{our,bliokh 2005,berard}, 
\beq \vec{R} = \vec{r} + \vec{{\cal A}}(\vec{p})\label{27},\eeq  $\vec{{\cal A}}(\vec{p})= \mu\frac{\vec{p}\times\vec{\sigma}}{p^{2}}$, where $\vec{r}(\vec{p})$ is the mean position (momentum) of the external observable space.

It is well known that the field of an electromagnetic plane wave with wave vector $k_{\mu}$ ($k^{2}$ = 0) depends on the $4$-coordinates only in the combination $kx = \zeta.$ So, we write the gauge potential of the external laser field as $A_{\mu}(\zeta).$ If $A_{\mu}(\zeta)$ are periodic functions and their time average value $\langle A_{\mu}(\zeta)\rangle = 0,$ the time average value of the modified momentum 4-vector of the electron $P^{\mu}$ in the laser field is given by \cite{book}
\beq P^{\mu} = p^{\mu} - \frac{e^{2}\langle A^{2}\rangle}{2(kp)}k^{\mu}\label{l9}.\eeq The spatial component of the momentum can now be written as 
\beq \vec{P} = \vec{p} - \frac{e^{2}\langle A^{2}\rangle}{2(kp)}\vec{k} = \vec{p} - \alpha\vec{k} = \vec{p} - \vec{k}^{~'}\label{20},\eeq
with $\alpha = \frac{e^{2}\langle A^{2}\rangle}{2(kp)}.$ If we now consider the situation of the EVB having head-on collision with the laser field, $\vec{k}$ is anti-parallel to $\vec{p}$ so that we can write the modified momentum as \beq \vec{P} =  \vec{p} + \vec{k}^{~'}.\label{21}\eeq 

 Thus from equations (\ref{27}) and (\ref{21}) we can write the angular momentum as
\begin{widetext}
\beq \vec{{\tilde L}} = \vec{R}\times \vec{P} = (\vec{r} + \vec{{\cal A}}(\vec{p}) )\times (\vec{p} + \vec{k}^{~'})= \vec{r} \times \vec{p} + \vec{r}\times \vec{k}^{~'} + \vec{{\cal A}}(\vec{p}) \times \vec{p} + \vec{{\cal A}}(\vec{p}) \times \vec{k}^{~'} = \vec{L}_{1}+ \vec{L}_{2} + \vec{L}_{3} + \vec{L}_{4}\label{28}.\eeq
\end{widetext}
In eqn. (\ref{28}), $\vec{L}_{1}$ represents the OAM of the field-free EVB and $\vec{L}_{2}$ corresponds to the additional angular momentum $n,$ which is induced by the laser field. Now to compute $\langle\vec{L}_{3} \rangle,$ we write \beq \vec{L}_{3} = \vec{{\cal A}}(\vec{p}) \times \vec{p}  = \mu \frac{\vec{p}\times \vec{\sigma}\times\vec{p}}{p^{2}} = -\mu \vec{p}\times \frac{\vec{p}\times \vec{\sigma}}{p^{2}}\label{29},\eeq where $\vec{\sigma}$ is the vector of Pauli matrices. The expectation value of $\vec{\sigma}$ is given by
\beq \langle\vec{\sigma}\rangle = \frac{\langle\psi|\vec{\sigma}|\psi\rangle}{\langle\psi|\psi\rangle} = \vec{n}^{~'},\label{30}\eeq where $\psi$ is a two-component spinor \beq \psi = \left(\begin{array}{cr}
   \psi_{1} \\
    \psi_{2}
 \end{array}\right) \label{31}\eeq and $\vec{n}^{~'}$ is the unit vector. Thus 
\beq \langle \vec{L}_{3}\rangle = \langle \mu\vec{p}\times(\frac{\vec{\sigma}\times\vec{p}}{p^{2}})\rangle = - \langle \mu \vec{\kappa}\times(\vec{\kappa}\times \vec{\sigma}) \rangle,\eeq
 with $\frac{\vec{p}}{p} = \vec{\kappa},$ $\vec{\kappa}$ being the unit vector. This gives  \beq \langle \vec{L}_{3}\rangle =-\mu\vec{n}^{~'}.\label{32m}\eeq 
Besides, as the momentum vectors $\vec{p}$ and $\vec{k}^{~'}$ are anti-parallel to each other,
$ \langle \vec{L}_{4}\rangle = \vec{k}^{~'} \times \mu(\frac{\vec{p}\times\vec{\sigma}}{p^{2}}) = 0 $ 

The first two terms in equation (\ref{28}) dictate the addition of OAM $n$ with the field-free OAM $l$ of EVB due to the laser field. This is caused by the orbit-orbit interaction between the intrinsic OAM and the OAM owing to the external degrees of freedom. The local vortex structure $exp(il\phi)$ in the field-free wave packet in the Bessel beam spectrum is now modified as  $exp(il^{'}\phi),$ where $l^{'}=l+n,$ giving rise to a magnetic monopole type of  Berry connection. In terms of $\vec{e},$  a unit vector orthogonal to $\vec{P}, $ we can write \beq exp(il^{'}\phi) = (e_{x} + ie_{y})^{l^{'}}\label{32}.\eeq 
It is noted that with the variation of $\vec{P},$ $\vec{e}$ moves on the unit sphere $\frac{\vec{P}}{|\vec{P}|},$ which leads to the monopole type connection \cite{bliokh 2007}
\beq \vec{{\cal A}} = \left(i (e_{x} - ie_{y})\frac{\partial}{\partial \vec{P}}(e_{x} + ie_{y})\right)\label{33}\eeq and the corresponding curvature $\vec{\Omega}(\vec{P}) = \frac{\vec{P}}{P^{3}}.$ As a result, we have $\vec{{\cal A}}^{(l^{'})} = l^{'}\vec{{\cal A}}$ and $\vec{\Omega}^{(l^{'})} = l^{'}\vec{\Omega}$ indicating that the charge of the magnetic monopole in momentum space is given by $l^{'}.$ Noting that the electric field component of the external field will accelerate electrons, the momentum concerned will be time dependent and lead to an anomalous velocity as 
\beq \vec{v}_{a} = l^{'}\dot {\vec{P}}\times \vec{\Omega}(\vec{P}) = l^{'}\dot {\vec{P}}\times \frac{\vec{P}}{P^{3}}.\label{34}\eeq This will give rise to OAM Hall effect \cite{bliokh 2007}. Thus, analogous to the spin-orbit coupling giving rise to the spin Hall effect, the orbit-orbit interaction between the intrinsic OAM and the external degrees of freedom gives rise to OAM Hall effect. In case of a paraxial beam, as mentioned earlier, the Berry phase factor $\mu$ vanishes and so there will be no contribution from $\langle \vec{L}_{3}\rangle$ here as follows from eqn. (\ref{32m}). This argument convincingly demonstrates that a spatial shift of the center of the EVB along the polarization direction of photons in the laser field is caused by the OAM Hall effect and the shift will depend on the orbital angular momentum. Large values of the orbital angular momentum can cause a larger amount of shift.   

For the analysis of the non-paraxial beams arising out of the tilted vortices equations (\ref{28}) and (\ref{32m}) are used and we write the total angular momentum of the EVB in presence of a laser field as 
$\langle \vec{{\tilde L}}\rangle = (l + n+ \mu) {\hat{\vec{z}}} = (l^{'} + \mu){\hat{\vec{z}}}.\label{35} $ 
From the conservation law of the total angular momentum  $\vec{{\tilde L}} + \vec{{\tilde S}} = \vec{L} + \vec{S},$ where $\vec{{\tilde S}}$ corresponds to the spin vector with $\vec{L} = l^{'}{\hat{\vec{z}}}$ and $\vec{S} = s{\hat{\vec{z}}},$ we find 
$ \langle \vec{{\tilde S}}\rangle =  (s - \mu){\hat{\vec{z}}}.\label{36}$ 
The presence of spin-orbit interaction (SOI) is implied due to the conversion of a part of the angular momentum from SAM to OAM.
The quantized value of $\mu = \frac{1}{2}$ corresponds to the relation $|\mu| = s,$ whereas for non-quantized value of $\mu,$ the expectation value $ \langle\vec{\tilde{L}}\rangle(\langle\vec{\tilde{S}}\rangle)$ can take arbitrary values. Indeed, from the relation of the angular momentum in presence of a magnetic monopole $\vec{J} = \vec{L} - \mu\hat{\vec{r}},$ it is noted that for vanishing $\vec{L},$ the total angular momentum is $\mu$ and for $\mu = \frac{1}{2}$ we have the intrinsic angular momentum of the system given by $\frac{1}{2},$ which is the SAM of an electron with $s_{z} = \pm \mu.$ 


The non-quantized value of $\mu$ undergoes renormalization group (RG) flow \cite{20,21} following the relation $L\frac{\partial \mu}{\partial L}\leq 0,$ where $L$ is a length scale. This suggests that for non-quantized values of $\mu$(denoted as ${\tilde\mu}$) we can consider it as a continuous function and at certain fixed points in the parameter space the monopole charge corresponds to  quantized values. 

We now introduce a non-inertial coordinate frame with basis vectors $(\vec{v}, \vec{w}, \vec{u})$ attached to the local direction of momentum $\vec{u} = \frac{\vec{P}}{|P|}.$ This coordinate frame rotates with the variation of $\vec{P}.$ With respect to a motionless (laboratory) coordinate frame such rotations describe a precession of the triad $(\vec{v}, \vec{w}, \vec{u}),$ with some angular velocity. At an instant of time, if we take the direction of the vortex line as the local $z$ axis which represents the direction of propagation of the wave front, this corresponds to the paraxial beam in the local frame. In this non-inertial local frame the local monopole charge will correspond to a pseudospin. In fact the expectation value of the spin operator
$ \left\langle \vec{S}\right\rangle = \frac{1}{2}\frac{\langle\psi|\vec{\sigma}|\psi\rangle}{\langle\psi|\psi\rangle}\label{43},$
undergoes precession with the precession of the coordinate frame \cite{bliokh 2009}. When the direction of the vortex line is taken to be the local $z$ axis, the local value of $\tilde{\mu}$ is changed and takes the quantized value $|\mu| = \frac{1}{2}$ owing to the precession of the spin vector and thus corresponds to the pseudospin  in this frame. The pseudospin vector $\vec{S}$ is parallel to the momentum vector $\vec{P}.$ We can now formulate an anomalous velocity as 
\beq \vec{v}_{a} = \mu \dot{\vec{P}}\times\frac{\vec{P}}{P^{3}} \label{37}.\eeq
Thus the anomalous velocity is perpendicular to the pseudospin vector and points along opposite directions depending on the chirality  $s_{z} = \pm \frac{1}{2}$ corresponding to $\mu> 0(< 0).$ This separation of the spins gives rise to the spin Hall effect. Thus a tilted vortex line with respect to the propagation direction in the inertial frame carrying OAM will give rise to spin Hall effect, which is caused by the spin-orbit interaction.

The expression for $v_{a}$ in equation (\ref{37}) can be rewritten in terms of the unit vector $\vec{u} = \frac{\vec{P}}{|\vec{P}|}$ and its time derivative as
\beq \vec{v}_{a} = \mu\frac{\dot{\vec{P}}\times\vec{P}}{P^{3}} = \mu \dot{\vec{u}}\times\vec{u}  .\label{38}\eeq Denoting $\frac{\dot{\vec{u}}}{|\dot{\vec{u}}|} = \vec{n}_{1},$ we note that the spin current is orthogonal to the local plane ($\vec{u}, \vec{n}_{1}$). Since the spin current is orthogonal to the local plane $(\vec{u},\vec{n}_{1})$, thus for a tilted vortex with respect to the wave propagation direction the spin Hall effect is a Coriolis type transverse deflection. This leads to a shift of the center of the beam with respect to the center of field-free EVB in case of non-paraxial beams.


In conclusion, we have considered the relativistic EVBs in a laser field and explored the  dynamics of the system  from the perspective of geometric phase. It has been argued that in case of the paraxial beams, a shift of the center of the beam in the polarization direction of the laser field is caused by OAM Hall effect whereas for non-paraxial beams the shift is a consequence of the spin Hall effect. The arbitrary large integer value of the angular orbital momentum suggests the generation of a larger shift in the paraxial case than that in the non-paraxial case which is in conformity with the numerical estimates given in the work of  Hayrapetyan et.al\cite{laser}. It may be added here, the shifts of the center of the beam with respect to the center of field-free EVB discussed in \cite{laser} have been estimated by  calculating the probability density of finding an electron in the beam profile. 
However, our  geometric phase inspired non-trivial analysis unveils the dynamics of the shift in terms 
of the anomalous velocity that is induced due to the associated Berry curvature in the Hall effect scenario. 

Acknowledgement: The authors appreciate the encouraging and helpful comments of the anonymous referees.

\end{document}